\documentclass{amsart}

\usepackage[latin1]{inputenc}
\usepackage[english]{babel}
\usepackage{tikz}
\usepackage{subfigure}
\usepackage[ruled,vlined,titlenumbered,linesnumbered,longend]{algorithm2e}
\usepackage{verbatim}

\title{How to generate an object under an ordinary Boltzmann distribution via an exponential Boltzmann sampler.}
\author[O. Bodini]{O. Bodini$^{\dagger}$}
\address{$\dagger$ olivier.bodini@lip6.fr}
\thanks{$\dagger$ ANR GAMMA 07-2\_195422 grant of the French \textit{Agence Nationale de la recherche.}}

\begin{document}

\maketitle

\begin{abstract}
This short note presents an efficient way to derive from an exponential Boltzmann sampler a ordinary Boltzmann sampler.\\
Cette note rapide présente une façon efficace de passer d'un générateur de Botzmann exponentiel à un générateur de Boltzmann ordinaire.  
\end{abstract}

\section*{Introduction and Motivations}
Les générateurs de Boltzmann introduit par Duchon, Flajolet, Louchard et Schaeffer \cite{bolt1}, constituent une avancée importante pour la génération aléatoire. Ils permettent aujourd'hui d'engendrer des objets de tailles inexpérées avec les techniques précédentes. Nous montrons dans cette petite note comment transformer un générateur exponentiel de Boltzmann pour une classe combinbatoire $\mathcal{C}$, c'est à dire un générateur dont la probabilité de tirer un objet $\omega$ de taille $n$ est $\frac{x^n}{n!\hat{C}(x)}$ en un générateur ordinaire de Boltzmann : i.e. avec la probabilité de tirer un objet $\omega$ de taille $n$ en $\frac{x^n}{C(x)}$. Ce petit résultat est le premier maillon nécessaire à l'élaboration d'un générateur de Boltzmann pour l'opérateur Shuffle \cite{DPRS}. 

\section{Notations et définitions de base}
 Nous renvoyons le lecteur au livre ``Analytic Combinatorics" \cite{livre} pour une remarquable exposition de ce qu'est la combinatoire symbolique. Cette note utilise la terminologie et les concepts de base de ce domaine des mathématiques. 

Nous rappelons ici brièvement les notions de combinatoire symbolique qui nous serons utiles par la suite :
Une \textit{classe combinatoire} est un couple $(\mathcal{A}, |.|)$ (simplement $\mathcal{A}$ s'il n'y a pas d'ambiguité) où $\mathcal{A}$ est l'ensemble des objets de la classe et $|.|$ une fonction taille $\mathcal{A} \rightarrow \mathbb{N}$ associé aux objets. De plus nous désirons qu'il n'y ait qu'un nombre fini d'objet de même taille.
Etant donnée $\mathcal{A}$ une classe combinatoire. Nous notons $\hat{A}(x)$  la \emph{serie génératrice exponentielle} de $\mathcal{A}$ (EGF): $\hat{A}(x) = \sum\limits_{n \geq 0} \frac{a_n}{n!} x^n $ où $a_n$ est le nombre d'objets dans $\mathcal{A}$ de taille $n$. Parallelement, la \emph{fonction génératrice ordinaire} de  $\mathcal{A}$  (OGF) est $A(x)= \sum\limits_{n \geq 0} {a}_n x^n $.\\

\section{Génération aléatoire}
On rappelle ici les deux types de générateurs de Boltzmann comme introduit dans \cite{bolt1}. Les \emph{générateurs exponentiels} qui sont des générateurs qui doivent renvoyer un objet $\omega$ de taille $n$ avec probabilité $\frac{x^n}{n!\hat{C}(x)}$ et les \emph{générateurs ordinaires} qui doivent renvoyer un objet $\omega$ de taille $n$ avec probabilité $\frac{x^n}{C(x)}.$

Montrons ci-dessous comment passer de manière efficace d'un générateur de Boltzmann exponentiel vers un générateur de Boltzmann ordinaire. L'idée est similliaire à celle présente dans l'article \cite{BRS,roussel} permettant de passer d'un générateur de Boltzmann d'une classe pointée $\mathcal{A}^\bullet$ à un générateur de Boltzmann pour la classe $\mathcal{A}$. Cette approche a été pour la première fois evoquée lors d'une discution entre P. Flajolet et M. Soria.
Nous reprenons donc ici l'idée de ne pas selectionner le parametre de Boltzmann de manière déterministe mais de le choisir de manière probabiliste à partir d'une distribution continue de probabilité bien choisie. Plus explicitement, en voici le principe :\\

\begin{algorithm}[H]
\caption{$\Gamma_x\mathcal{A}$}
\label{gammaBCI}

\KwOut{un objet tiré suivant une probabilité de Boltzmann ordinaire}
Tirer un parametre $u$ suivant la densité de probabilité $\dfrac{e^{-u}\hat{A}(xu)}{A(x)} $\\
Tirer un objet $\gamma$ avec le générateur exponentiel de paramêtre $xu$ : $\gamma=\hat{\Gamma}_{xu}\mathcal{A}.$

\Return l'objet $\gamma$.\\
\end{algorithm}

\begin{proof}
Tout d'abord $d(u)=\dfrac{e^{-u}\hat{A}(xu)}{A(x)}$ est bien une densité de probabilité sur $[0,\infty]$. C'est une fonction positive et $\displaystyle{\int\limits_{u=0}^\infty\dfrac{e^{-u}\hat{A}(xu)}{A(x)}du=1} .$
Il reste à montrer qu'un objet $\gamma$ de taille $n$ est bien retourné avec probabilité $\frac{x^n}{C(x)}.$ Pourque l'objet $\gamma$ soit retourné, il faut avoir tirer une certaine valeur pour $u$, puis avoir tiré $\gamma$ avec le générateur $\hat{\Gamma}_{xu}\mathcal{A}.$ Ceci donne la probabilité : $p(\gamma)=\displaystyle{\int\limits_{u=0}^\infty\dfrac{e^{-u}\hat{A}(xu)}{A(x)} \frac{(xu)^n}{n!\hat{A}(xu)}du}.$ Soit après symplification, $p(\gamma)=\displaystyle{\frac{x^n}{n!A(x)}\int\limits_{u=0}^\infty e^{-u} u^ndu}.$ Mais $\displaystyle{\int\limits_{u=0}^\infty e^{-u} u^ndu= n!}$ et le théorème s'ensuit.
\end{proof}

En conclusion, en utilisant des techniques classiques permettant de tirer suivant une certaine densité de probabilitié (voir L. Devroye), il est possible de faire un tirage de $u$ en temps constant et donc, l'algorithme ci-dessus et un générateur ordinaire de même complexité que le générateur exponentiel.

\section{remerciement}
En tout premier lieu, à P. Flajolet sans qui cette note n'aurait pas lieu.
\bibliographystyle{plain}
\bibliography{biblio}

\end{document}